\documentclass[aps,prd]{revtex4}% Physical Review D

\usepackage{graphicx}% Include figure files
\usepackage{dcolumn}% Align table columns on decimal point
\usepackage{bm}% bold math
\reversemarginpar
\usepackage{graphicx}
\usepackage{amsmath}
\def\a{\alpha}

\def\g{\gamma}
\def\d{\delta}

\def\k{\kappa}

\def\m{\mu}

\def\D{\Delta}

\def\beq{\begin{equation}}         \def\eeq{\end{equation}}
\def\beqa{\begin{eqnarray}}              \def\eeqa{\end{eqnarray}}
\def\be{\begin{eqnarray}}                 \def\ee{\end{eqnarray}}

\def\ba{\begin{array}}        \def\ea{\end{array}}
\def\bi{\begin{itemize}}        \def\ei{\end{itemize}}

\def\nn{\nonumber \\ }

\def\exp#1{e^{#1}}                    % exponent %
\def\float#1(#2){#1\times 10^{#2}}  % expression of float number%
\def\err#1(#2){#1 \pm #2}  % expression of experimental error%

\def\order#1{\ensuremath{{\cal O}(#1)} }% order of magnitude %

\def\bar#1{\overline{#1}}  % conjugate fields %
\def\bra#1{\ensuremath{ \left\langle #1 \right| } }  % quantum state %
\def\ket#1{\ensuremath{ \left| #1 \right\rangle} }  % conjugate quantum state %

\def\NPB#1#2#3    { Nucl. Phys. {\bf B#1}, #2 (#3)}
\def\npps#1#2#3   { Nucl. Phys. Proc. Suppl. {\bf #1}, #2 (#3)}
\def\PLB#1#2#3    { Phys. Lett. {\bf B#1}, #2 (#3)}
\def\PRD#1#2#3    { Phys. Rev. {\bf D#1}, #2 (#3)}
\def\prep#1#2#3   { Phys.Rep. {\bf #1} ,#2 (#3)}
\def\PRL#1#2#3    {{ Phys.~Rev.~Lett. }~{\bf #1} ,~#2~(#3)}
\def\ijm#1#2#3    { Int. j. Mod. Phys.{\bf A#1}  ,#2 (#3)}
\def\mpla#1#2#3   { Mod. Phys. Lett. {\bf A#1} ,#2 (#3)}
\def\zpc#1#2#3    { Zeit. f{\"u}r Physik {\bf C#1} ,#2 (#3)}

\def\Npb#1#2#3    { Nucl. Phys. {\bf B#1}, #3 (#2)}
\def\Npps#1#2#3   { Nucl. Phys. Proc. Suppl. {\bf #1}, #3 (#2)}
\def\Plb#1#2#3    { Phys. Lett. {\bf B#1}, #3 (#2)}
\def\Prd#1#2#3    { Phys. Rev. {\bf D#1}, #3 (#2)}
\def\Prep#1#2#3   { Phys. Rep. {\bf #1} ,#3 (#2)}
\def\Prl#1#2#3    { Phys. Rev. Lett. {\bf #1} ,#3 (#2)}
\def\Ijm#1#2#3    { Int. j. Mod. Phys.{\bf A#1}  ,#3 (#2)}
\def\Mpla#1#2#3   { Mod. Phys. Lett. {\bf A#1} ,#3 (#2)}
\def\Zpc#1#2#3    { Zeit. f{\"u}r Physik {\bf C#1} ,#3 (#2)}

\begin{document}

%============ The new commands: =============================
    \newcommand{\DSC}{D\hspace{-0.25cm}\slash_{\bot}}
    \newcommand{\DSP}{D\hspace{-0.25cm}\slash_{\|}}
    \newcommand{\DS}{D\hspace{-0.25cm}\slash}
    \newcommand{\DC}{D_{\bot}}
    \newcommand{\DSCX}{D\hspace{-0.20cm}\slash_{\bot}}
    \newcommand{\DSPX}{D\hspace{-0.20cm}\slash_{\|}}
    \newcommand{\DP}{D_{\|}}
    \newcommand{\QV}{Q_v^{+}}
    \newcommand{\QVB}{\bar{Q}_v^{+}}
    \newcommand{\QVP}{Q^{\prime +}_{v^{\prime}} }
    \newcommand{\QVBP}{\bar{Q}^{\prime +}_{v^{\prime}} }
    \newcommand{\QVHZ}{\hat{Q}^{+}_v}
    \newcommand{\QVHZB}{\bar{\hat{Q}}_v{\vspace{-0.3cm}\hspace{-0.2cm}{^{+}} } }
    \newcommand{\QVPHZB}{\bar{\hat{Q}}_{v^{\prime}}{\vspace{-0.3cm}\hspace{-0.2cm}{^{\prime +}}} }
    \newcommand{\QVPHFB}{\bar{\hat{Q}}_{v^{\prime}}{\vspace{-0.3cm}\hspace{-0.2cm}{^{\prime -}} } }
    \newcommand{\QVPHB}{\bar{\hat{Q}}_{v^{\prime}}{\vspace{-0.3cm}\hspace{-0.2cm}{^{\prime}} }   }
    \newcommand{\QVHF}{\hat{Q}^{-}_v}
    \newcommand{\QVHFB}{\bar{\hat{Q}}_v{\vspace{-0.3cm}\hspace{-0.2cm}{^{-}} }}
    \newcommand{\QVH}{\hat{Q}_v}
    \newcommand{\QVHB}{\bar{\hat{Q}}_v}
    \newcommand{\VS}{v\hspace{-0.2cm}\slash}
    \newcommand{\MQ}{m_{Q}}
    \newcommand{\MQP}{m_{Q^{\prime}}}
    \newcommand{\QVHPMB}{\bar{\hat{Q}}_v{\vspace{-0.3cm}\hspace{-0.2cm}{^{\pm}} }}
    \newcommand{\QVHMPB}{\bar{\hat{Q}}_v{\vspace{-0.3cm}\hspace{-0.2cm}{^{\mp}} }  }
    \newcommand{\QVHPM}{\hat{Q}^{\pm}_v}
    \newcommand{\QVHMP}{\hat{Q}^{\mp}_v}
 %==============================================================

%\newcommand{\be}{\begin{eqnarray}}
%\newcommand{\beq}{\begin{eqnarray}}
\newcommand{\befg}{\begin{figure}}
\newcommand{\CDP}{\hat{\cal D}\hspace{-0.27cm}\slash_v}
\newcommand{\CDPL}{\overleftarrow{\hat{\cal D}}\hspace{-0.29cm}\slash_v}
\newcommand{\CDPN}{{\cal D}\hspace{-0.26cm}\slash_v}
\newcommand{\CDPNL}{\overleftarrow{\cal D}\hspace{-0.29cm}\slash_v}
\newcommand{\Dbot}{{/\!\!\!\!D}\hspace{-3pt}_{\bot}}
\newcommand{\dsp}{D\hspace{-0.27cm}\slash_{\|}}
\newcommand{\DSCXL}{\overleftarrow{D}\hspace{-0.29cm}\slash_{\bot}}
\newcommand{\Dslashbot}{{/\!\!\!\!D}\hspace{-3pt}_{\bot}}
\newcommand{\Dslash}{/\!\!\!\!D}
\newcommand{\edfg}{\end{figure}}
\def\fvcb{${\cal F}(1)|V_{\rm cb}|$}
\newcommand{\kslash}{\not\!{k}}
\newcommand{\kslashbot}{{/\!\!\!\!k}\hspace{-3pt}_{\bot}}
\newcommand{\mbh}{\hat{m}_b}
\newcommand{\non}{\nonumber \\}
\def\Ova{O^q_{V-A}}
\def\Osp{O^q_{S-P}}
\newcommand{\oDbot}{\overleftarrow{/\!\!\!\!D}\hspace{-5pt}_{\bot}}
\newcommand{\oDSP}{\not\!{v} v\cdot \overleftarrow{D}}
\newcommand{\odsp}{\overleftarrow{D}\hspace{-0.29cm}\slash_{\|}}
\newcommand{\oDslashbot}{\overleftarrow{/\!\!\!\!D}\hspace{-5pt}_{\bot}}
\newcommand{\omiga}{\omega}
\newcommand{\pbot}{\partial\hspace{-3pt}_\bot}
\newcommand{\PP}{{1 + v\hspace{-0.2cm}\slash \over 2}}
\newcommand{\PM}{{1 - v\hspace{-0.2cm}\slash \over 2}}
\newcommand{\PSC}{\partial\hspace{-0.2cm}\slash_{\bot}}
\newcommand{\PSP}{\partial\hspace{-0.2cm}\slash_{\|}}
\newcommand{\ptslashbot}{\partial\hspace{--0.2cm}\slash_\bot}
\newcommand{\qpv}{Q_v^{(+)}}
\newcommand{\qpvb}{\bar{Q}_v^{(+)}}
\newcommand{\qslash}{\not\!{q}}
\newcommand{\uv}{/\!\!\!\!\hspace{1pt}v}
\newcommand{\uvslash}{/\!\!\!\!\hspace{1pt}v}
\newcommand{\vslash}{\not\!{v}}

%=========================================================
\title{ MORE PRECISE DETERMINATION of $V_{cb}$ $\&$
$V_{ub}$ and \\ DIRECT CP VIOLATION IN CHARMLESS B DECAYS
\footnote{Talk delivered by
Y. L. Wu at ICFP2003, KIAS, Korea.}}% Force line breaks with \\
 \author{Wen-Yu Wang} \email[Email: ]{wangwy@itp.ac.cn} %
\author{Yue-Liang Wu} \email[Email: ]{ylwu@itp.ac.cn}
 %\email{ylwu@itp.ac.cn}
 \affiliation{Institute of Theoretical Physics, Chinese Academy of sciences,
 Beijing 100080, China  }
 \author{Yu-Feng Zhou} \email[Email: ]{zhou@theorie.physik.uni-muenchen.de}
\affiliation{Ludwig-Maximilians-University Munich,  Sektion
Physik. Theresienstra$\beta$e 37, D-80333. Munich,Germany}
 %
%\author{ author for second address}
%\affiliation{
\date{November 18, 2003}
\begin{abstract}
More precise extractions on the two important CKM matrix elements
$|V_{cb}|$ and $|V_{ub}|$ are presented up to $1/m_Q^2$
corrections based on the complete heavy quark effective field
theory (HQEFT) of QCD. The HQEFT as a large component QCD provides
us a complete theoretical framework for correctly and
systematically evaluating the subleading and higher order
contributions in $1/m_Q$ expansion of heavy quarks. A global
analysis on charmless B decays is made based on the isospin and
SU(3) flavor symmetry. An isospin relation is found to be very
useful for studying SU(3) symmetry breaking effects of strong
phases and exploring new type of electroweak penguin effects. The
direct CP violation in charmless B decays is predicted and its
precise measurement is helpful either for testing SU(3) symmetry
breaking effects of strong phases or for probing new physics
beyond standard model.
\end{abstract}
%\pacs{ }%PACS numbers


\maketitle

\renewcommand{\theequation}%
 {\arabic{equation}}

\vspace{1cm}

{\bf Introduction}. The standard model (SM) has been well tested
in the gauge sector. The unsolved and unclear problems in the SM
are mainly concerned in the Yukawa sector which is strongly
related to flavor physics, such as: origin and mechanism of CP
violation, origins of the quark and lepton masses as well as their
mixing. It involves thirteen parameters whose origins are all
unknown. Therefore, precisely extracting those parameters and
testing CP violation mechanism as well as probing new physics
become hot topics in flavor physics. In fact, flavor physics has
already indicated the existence of new physics\cite{HF}. Much
efforts have been made by experimentalists in flavor
physics\cite{KK}. For instance, after forty years of discovery for
indirect CP violation, direct CP violation in kaon decays has
recently been established by two experimental
groups\cite{NA48,KTeV}, which is also consistent with the
theoretical predictions\cite{ylw}. CP violation has also been
observed in B decays at the two B-factories. More precise
experimental results in flavor physics will become available in
the recent years. It is known that exclusive semileptonic and
inclusive B decays play a crucial role for extracting two
important parameters $V_{cb}$ and $V_{ub}$ in the CKM matrix
elements. Rare B decays and direct CP violations are also of great
importance in determining weak phase angles of the unitarity
triangle and testing the Kobayashi-Maskawa (KM) mechanism
~\cite{kobayashi:1973fv} in SM as well as probing new physics. In
this talk, we will pay attention to the more precise extraction of
the CKM matrix elements $V_{cb}$ and $V_{ub}$ and the direct CP
violation in charmless B decays.

{\bf Complete HQEFT of QCD and incompleteness of the usual HQET}.
Heavy quark effective field theory (HQEFT) of QCD, which was first
explored in\cite{W0} and recently developed in detail by a series
of papers\cite{W1,W2,W3,W4,W5,W6,W7,W8,W9,W10,W11}, provides a
promising and systematic tool in correctly evaluating the hadronic
matrix elements of heavy quarks and precisely extracting the CKM
matrix elements $V_{cb}$ and $V_{ub}$ from B decays via heavy
quark expansion (HQE). As the HQEFT is a theoretical framework
derived directly from QCD, it explicitly displays the heavy quark
symmetry (HQS)\cite{HQS} in the infinite mass limit
$m_Q\rightarrow \infty$\cite{HQL} and symmetry breaking
corrections for finite mass case in the real world. In fact, the
HQEFT has been shown to be as a large component QCD\cite{W0,W10}.
At the leading order, it coincides with the usual heavy quark
effective theory(HQET)\cite{HQET} which is constructed based on
the heavy quark symmetry in the infinite mass limit. The
differences between HQEFT of QCD and the usual HQET arise from the
subleading terms in the $1/m_Q$ expansion. This is because in the
construction of HQET the particle and antiparticle components were
separately treated based on the assumption that the particle
number and antiparticle number are conserved separately in the
effective Lagrangian. However, such an assumption is only valid in
the infinite mass limit. Note that the particle number and
antiparticle number are always conserved in the transition matrix
though the particle and antiparticle number is not conserved in
the Lagrangian due to the pair creation and annihilation
interaction terms, which is independent of heavy quark limit and
is in fact the basic principle of quantum field theory. Obviously,
the quark-antiquark coupled terms that correspond to the pair
creation and annihilation interaction terms in full QCD were
inappropriately dropped away in the usual HQET. Those terms have
been shown in HQEFT of QCD to be suppressed by $1/m_Q$ and they
truly become vanishing in infinite mass limit. Thus the usual HQET
based on the assumption of particle and antiparticle number
conservation in the effective Lagrangian cannot be regarded as a
complete effective theory for evaluating the subleading and higher
order corrections, it must be an incomplete effective theory.
Unlike the derivation of the usual HQET from QCD by making the
assumption of particle and antiparticle number conservation in the
effective Lagrangian\cite{MR}, we have derived the HQEFT \cite{W0}
from full QCD by carefully treating all contributions of the field
components, i.e., large and small, `particle' and `antiparticle'
in the effective Lagrangian, so that the resulting effective
Lagrangian should form the right basis for a {\bf complete}
effective field theory of heavy quarks\cite{W10}. Though the
quark-antiquark mixing terms are suppressed by $1/m_Q$, their
physics effects at subleading and higher orders have been found in
some cases to be significant and crucial for obtaining consistent
results \cite{W1,W2,W3,W4,W5,W6,W7,W8,W9}. Therefore, to correctly
and precisely consider the finite quark mass corrections, it is
necessary to include the contributions from the components of the
antiquark fields. Namely one must integrate in the contributions
of the antiquark (quark) field to the effective lagrangian and the
transition matrix element when considering the quark (antiquark)
processes, which is equivalent to consider all the tree diagram
contributions via virtual antiquark (quark) exchanges. As the
quark and antiquark interaction terms are going to be decoupled
only in the infinite mass limit, the treatment of the usual HQET
for subleading and higher order corrections is incomplete though
it has been widely applied to various processes\cite{HQA}. Before
proceeding, we would like to address the following points that may
be misunderstood in some literatures. Firstly, the Wilson
coefficient functions mainly characterize the perturbative effects
calculated in full QCD, the nonperturbative effects below the
energy scale $m_Q$ are described by the HQEFT of QCD via the
$1/m_Q$ expansion, so that the antiquark effects in HQEFT of QCD
cannot be attributed to the Wilson coefficient functions.
Secondly, as the antiquark effects appear at the subleading order
in $1/m_Q$ expansion, it should not surprise that at the leading
order of $1/m_Q$ expansion, the resulting anomalous dimensions
from both full QCD and HQEFT of QCD (or HQET) are all the same,
which must easily be understood from the matching requirement.
Clearly, the perturbative QCD expansion in $\alpha_s$ and the
heavy quark expansion in $1/m_Q$ are two different expansions.

{\bf More Precise Extraction of $|V_{cb}|$}. {\bf Exclusive
semileptonic decays} $B\to D^*(D)l\nu$ provide one of the main
approaches to extract $|V_{cb}|$. The differential decay rates
are:
   \begin{eqnarray}
   \label{widthb2ds}
  \frac{d\Gamma(B\rightarrow D^{\ast}l\nu)}{d\omega} &=&\frac{G^2_F}{48\pi^3}(m_B-m_{D^{\ast}})^2
    m^3_{D^{\ast}}\sqrt{\omega^2-1}(\omega+1)^2 \nonumber\\
    &&  \times  [1+\frac{4\omega}{\omega+1}
    \frac{m^2_B-2\omega m_B m_{D^{\ast}}+m^2_{D^{\ast}}}
    {(m_B-m_{D^{\ast}})^2}]\vert V_{cb} \vert^2 {\cal F}^2(\omega) ,\\
\label{widthb2d}
    \frac{d\Gamma(B\rightarrow Dl\nu)}{d\omega}&=&\frac{G^2_F}{48\pi^3}(m_B+m_{D})^2
    m^3_{D}(\omega^2-1)^{3/2} \vert V_{cb} \vert^2 {\cal G}^2(\omega)
   \end{eqnarray}
   with
\begin{eqnarray}
   \label{zerorecoilF}
     {\cal F}(1) =
%     \eta_{QED}
     \eta_{A} h_{A_1}(1) =\eta_{A} (1+\delta^*),\quad
  % \label{zerorecoilG}
     {\cal G}(1) = \eta_{V} [h_{+}(1)-\frac{m_B-m_D}{m_B+m_D}
      h_{-}(1)]
      =  \eta_{V} (1+\delta ) ,
\end{eqnarray}
where the QCD radiative corrections to two loops give the short
distance coefficients $ \eta_A=0.960\pm 0.007 $ and
$\eta_V=1.022\pm 0.004$ \cite{ac}. $\omega$ is the product of the
four-velocities of the $B$ and $D^* (D) $ mesons, $\omega=v\cdot
v'$. The weak transition form factors $h_{A_1}(\omega)$,
$h_+(\omega)$ and $h_-(\omega)$ can be expanded in powers of
$1/m_Q$ and represented by the heavy quark spin-flavor independent
wave functions. Up to order $1/m^2_Q$ and neglecting the
contributions of operators containing two gluon field strength
tensors, one has at the zero recoil point $\omega=1$
\begin{eqnarray}
\label{formfactorha1}
    h_{A_{1}}&=&1+\frac{1}{8\bar{\Lambda}^2}[\frac{1}{m_b}(\kappa_1+3\kappa_2)
     -\frac{1}{m_c}(\kappa_1-\kappa_2)]^2
     -\frac{1}{8m^2_b\bar{\Lambda}^2}
     (F_{1}+3F_{2}
     -2\bar{\Lambda}\varrho_{1}-6\bar{\Lambda}\varrho_{2})  \nonumber\\
    & & -\frac{1}{8m_c^2\bar{\Lambda}^2}(F_{1}-F_{2}-2\bar{\Lambda}\varrho_{1}
     +2\bar{\Lambda}\varrho_{2})  +\frac{1}{4m_b m_c
\bar{\Lambda}^2}(F_{1}+F_{2}-2\bar{\Lambda}\varrho_{1}
     -2\bar{\Lambda}\varrho_{2}) , \\
    \label{formfactorhz}
h_{+}&=&
1+\frac{1}{8\bar{\Lambda}^2}(\frac{1}{m_b}-\frac{1}{m_c})^2
     [(\kappa_1+3\kappa_2)^2-(F_1+3F_2) +2\bar{\Lambda}(\varrho_1+3\varrho_2)] ,
     \qquad \quad
%\label{formfactorhf}
h_{-} = 0,
    \end{eqnarray}
where the parameters in the rhs. of
Eqs.(\ref{formfactorha1})-(\ref{formfactorhz}) are defined in
HQEFT \cite{W1}. The binding energy of a heavy meson $M$ is
defined as $\bar{\Lambda} \equiv \lim_{m_Q \to \infty}
\bar{\Lambda}_M =\lim_{m_Q \to \infty} (m_M-m_Q)$. It is seen from
Eqs.(\ref{formfactorha1}) and (\ref{formfactorhz}) that both
$h_{A_1}$ and $h_+$ in HQEFT of QCD do not receive $1/m_Q$ order
correction at the zero recoil point. More interestingly, the
vanishing value of $h_-$ in HQEFT  arises from the fact of partial
cancellation between the $1/m_Q$ correction in the current
expansion and the $1/m_Q$ correction coming from the insertion of
the effective Lagrangian into the transition matrix elements
\cite{W1}.  Such a cancellation is not observed in the incomplete
HQET because in the latter framework the quark-antiquark couplings
are not taken into account explicitly. Though the HQET protects
the weak transition matrix elements from $1/m_Q$ order correction
at zero recoil point, it does not protect $h_-$ from such
correction\cite{am}. As a result, in the incomplete HQET
framework, only the $B\to D^* l\nu$ decay rate at zero recoil is
strictly protected against $1/m_Q$ order correction while $B\to
Dl\nu$ decay rate is not. This is the main reason besides the
experimental considerations for the conclusion that the $B\to
Dl\nu$ decay is not as favorable as $B\to D^*l\nu$ decay for
$|V_{cb}|$ extraction in HQET.
While the complete HQEFT of QCD is more reliable to extract
$|V_{cb}|$ from both $B\to D^* l\nu$ and $B\to Dl\nu$ decays,
because both rates of these decays do not receive $1/m_Q$ order
correction, as can be seen from
Eqs.(\ref{widthb2ds})-(\ref{formfactorhz}).

Furthermore, the HQEFT of QCD gives interesting relations between
meson masses and wave functions \cite{W1}, which enable one to get
the zero recoil values of HQEFT wave functions straightforwardly
from the heavy meson mass spectrum. On the other hand, the wave
functions can also be calculated via other nonperturbative
approaches such as sum rules. As these parameters are estimated
consistently, $|V_{cb}|$ can be determined to a good accuracy.

%=============================================

{\bf Inclusive semileptonic $B$ decays} are the other alternative
to determine $|V_{cb}|$. In the usual HQET the light quark in a
hadron is generally treated as a spectator, which does not affect
the heavy hadron properties to a large extent. This treatment may
be one possible reason for the failure of HQET in some
applications. For example, the world average value for bottom
hadron lifetime ratio $\tau(\Lambda_b)/\tau(B^0)$ can not be
explained well in the usual framework of HQET \cite{mc,mjch}.
Instead of simply applying the equation of motion for infinitely
heavy free quark, $iv\cdot D \QV=0$, we treat the heavy quark in a
hadron as a dressed particle, which means that the residual
momentum $k$ of the heavy quark within a hadron is considered to
comprise contributions from the light degrees of freedom. This
simple picture is adopted to conveniently take into account the
effects of light degrees of freedom and the binding effects of
heavy and light components of the hadron but not deal with the
complex dynamics of hadronization directly. Explicitly, one may
use the relation $\langle iv\cdot D \rangle \equiv \langle
H_v|\QVB iv\cdot D \QV|H_v\rangle /2\bar{\Lambda}_H \approx
\bar{\Lambda} \ne 0$. To acquire a good convergence of HQE, we
perform the expansion in terms of $k-v \langle iv\cdot D \rangle$
(or say, equivalently, in terms of $1/(m_Q+\bar{\Lambda}))$. Then
the $B\to X_cl\nu$ decay rate is found to be \cite{W2,W4}
\begin{eqnarray}
\label{incwidth}
 \Gamma(B\to X_c l \nu)=\frac{G^2_F \hat{m}^5_b V^2_{cb}}{192\pi^3}
   \eta_{cl}(\rho,\rho_l,\mu) \{ I_0(\rho,\rho_l,\hat{\rho})+I_1(\rho,\rho_l,\hat{\rho})
\frac{\kappa_1}{3 \hat{m}^2_b}
-I_2(\rho,\rho_l,\hat{\rho})\frac{\kappa_2}{\hat{m}^2_b} \},
\end{eqnarray}
where $\hat{m}_b=m_b+\bar{\Lambda}$, $I_0$, $I_1$ and $I_2$ are
functions of the mass square ratios $\rho={m^2_c}/{\hat{m}^2_b}$,
$\hat{\rho}^2={\hat{m}^2_c}/{\hat{m}^2_b}$ and
$\rho^2_l={m^2_l}/{\hat{m}^2_b}$. Here the calculation is
performed up to nonperturbative order $1/\hat{m}^2_Q$ and
perturbative order $\alpha^2_s$. The function $\eta_{cl}$
characterizes QCD radiative corrections \cite{mmm,mmmb}.
$\kappa_2$ is often extracted from the known $B-B^*$ mass
splitting $\kappa_2 \simeq \frac{1}{8} (m^2_{B^{*0}}-m^2_{B^0}
)\approx 0.06 \mbox{GeV}^2 $ which is consistent with the sum rule
result \cite{W3}. There are several points to be mentioned for
Eq.(\ref{incwidth}). Firstly, in deriving Eq.(\ref{incwidth}) the
effects of light degrees of freedom are explicitly accounted for
in the picture of a dressed heavy quark in a hadron. Secondly, it
is seen that the next leading order contributions vanish in our
HQE in terms of the inverse dressed heavy quark mass,
$1/\hat{m}_b$. Furthermore, our HQE in terms of $k-v \langle
iv\cdot D \rangle$ (or $1/\hat{m}_b$) has a good convergence. It
is found that the $1/\hat{m}^2_b$ order contributions induce only
$-0.7\sim 5\%$ corrections to the total width $\Gamma(H_b \to X_c
e \bar{\nu})$. Therefore we conclude that the higher order
nonperturbative corrections can be safely neglected. Finally, now
one needs only to treat the dressed quark mass
$\hat{m}_b=m_b+\bar{\Lambda}$ instead of considering the
uncertainties arising from the two quantities $m_b$ and
$\bar{\Lambda}$ separately. Note that these features can not be
observed in the HQE in the usual HQET, where one assumes $\langle
iv\cdot D \rangle $ to be zero or of higher order of $1/m_b$. In
HQET the next leading order corrections can be absent only when
the HQE is performed in terms of $1/m_b$, and the heavy quark mass
$m_b$ and the binding energy $\bar{\Lambda}$ have to be treated
separately. Thus in HQET the theoretical prediction of the total
decay width strongly depends on the value of bottom quark mass
$m_b$.

 {\bf Numerical Values of $V_{cb}$}. Using the current world average \cite{PDG} $ | V_{cb} | {\cal
F}(1)=0.0383\pm 0.0005 \pm 0.0009$ , $ |V_{cb} |{\cal
G}(1)=0.0413\pm 0.0029 \pm 0.0027 $, the lifetime
$\tau(B^0)=1.540\pm 0.014 \mbox{ps}$ and the branching ratio from
CLEO $\mbox{Br}(B\to X_c e \nu)=(10.49\pm 0.17 \pm 0.43 ) \% $, we
have\cite{W11}
\begin{eqnarray}
& & |V_{cb}|=0.0395 \pm 0.0011_{\mbox{exp}}\pm 0.0019_{\mbox{th}} \qquad \mbox{from} \quad  B\to D^*l\nu ,
\quad O(1/m^2_Q) \nonumber \\
& & |V_{cb}|=0.0434 \pm 0.0041_{\mbox{exp}}\pm 0.0020_{\mbox{th}} \qquad \mbox{from} \quad B\to Dl\nu ,
\quad O(1/m^2_Q) \\
& &|V_{cb}|=0.0394 \pm 0.0010_{\mbox{exp}}\pm 0.0014_{\mbox{th}}
\qquad \mbox{from} \quad B\to X_c l\nu, \quad O(1/m^2_Q) \nonumber
\end{eqnarray}
where the result extracted from $B\to D l\nu$ decay receives a
larger experimental uncertainty than that from $B\to D^* l\nu $
decay but a similar theoretical uncertainty as the latter. The
result obtained from $B\to D^*l\nu$ decay agrees quite well with
that from inclusive $B\to X_c l\nu$ decay.
These results then give the average\cite{W11}
\begin{equation}
 \label{vcbaverage}
 |V_{cb}|=0.0402 \pm 0.0014_{\mbox{exp}} \pm 0.0017_{\mbox{th}}.
\end{equation}
Alternatively it can be represented as
$
 A=0.83 \pm 0.07
$ in the Wolfenstein parameterization\cite{LW} $|V_{cb}|=A
\lambda^2$ with $\lambda=|V_{us}|=0.22$.

%===========================================

{\bf More precise extraction of $|V_{ub}|$}. It is similar to that
of $|V_{cb}|$. The main difference is now one generally has to
deal with heavy-to-light decays. For exclusive decays $B\to \pi
(\rho) l\nu$, one can parameterize the leading order transition
matrix elements in HQEFT as \cite{gzmy,hly}
\begin{eqnarray}
 \langle \pi(p)|\bar{u} \Gamma \QV|B_v\rangle  = -Tr[\pi(v,p)\Gamma {\cal
 M}(v)],\quad
\langle \rho(p,\epsilon^*)|\bar{u} \Gamma \QV|B_v\rangle
   = -i Tr[\Omega(v,p)\Gamma {\cal M}_v]   ,\nonumber
\end{eqnarray}
where
\begin{eqnarray}
  \pi(v,p)&=&  \gamma^5 [A(v\cdot p,\mu)+ {\hat{p}\hspace{-0.17cm}\slash}
 B(v\cdot p,\mu)], \nonumber\\
  \Omega(v,p)&=&    L_1(v\cdot p) {\epsilon\hspace{-0.17cm}\slash}^*
   +L_2( v\cdot p) (v\cdot \epsilon^*)
  + [L_3(v\cdot p)
   {\epsilon\hspace{-0.17cm}\slash}^* +L_4(v\cdot p) (v\cdot \epsilon^* )]
   {\hat{p}\hspace{-0.17cm}\slash}
\end{eqnarray}
with $\hat{p}^\mu=\frac{p^\mu}{v\cdot p}$. A, B and
$L_i(i=1,2,3,4)$ are the leading order wave functions in HQEFT.
HQS and relevant effective theories are useful for studying
heavy-to-light decays as they give us relations between different
channels. For example, since $L_i$ are heavy quark mass
independent, $B\to \rho l\nu$ and $D\to \rho l\nu$ are
characterized by the same set of wave functions $L_i$. In this
sense, HQS and effective theories simplify the heavy-to-light
decays, though for a single channel the number of independent
functions is not reduced. We calculated these wave functions from
light cone sum rules, considering the $\pi$ distribution functions
up to twist 4 and the $\rho$ distribution functions up to twist 2
for $B\to \pi l\nu$ decay \cite{W5} and for $B\to \rho l\nu$ decay
\cite{W6}. Recently the calculation on $B\to \pi l\nu$ up to
$1/m_Q$ order has also been performed \cite{W9} and the finite
mass correction in HQEFT is found to be small. Inclusive decays
$B\to ul\nu$ can be investigated in a similar way as for $B\to
cl\nu$ decays by using the HQE in HQEFT.

 {\bf Numerical Values of $V_{ub}$}. Our numerical results are summarized as follows
\begin{eqnarray}
& &  |V_{ub}|_{LO}=(3.4\pm 0.5_{\mbox{exp}} \pm 0.5_{\mbox{th}})
   \times 10^{-3}  \qquad \mbox{from} \quad B\to \pi l\nu  , \nonumber \\
& & |V_{ub}|_{LO}=(3.7\pm 0.6_{\mbox{exp}} \pm 0.7_{\mbox{th}}
)\times 10^{-3} \qquad \mbox{from} \quad B\to \rho l\nu , \nonumber \\
& & |V_{ub}|_{NLO}= (3.2 \pm 0.5_{\mbox{exp}} \pm
0.2_{\mbox{th}})\times 10^{-3}
\qquad \mbox{from} \quad  B\to \pi l\nu, \\
& &  |V_{ub}|=(3.48 \pm 0.62_{\mbox{exp}}\pm
0.11_{\mbox{th}})\times 10^{-3}  \qquad \mbox{from} \quad  B\to
X_u l\nu  \quad O(1/m^2_Q) \nonumber
\end{eqnarray}
All these results can be compared with the average of CLEO
\cite{PDG}: $|V_{ub}|=(3.25^{+0.25}_{-0.32}\pm 0.55)\times
10^{-3}$.

%-----------------------------------------------------------------
 {\bf Summary I}. Within the framework of complete HQEFT of QCD,
 various processes and approaches all lead to consistent and more precise values for $|V_{cb}|$ and
$|V_{ub}|$.
%========================================================

{\bf Charmless B decays and direct CP violation}.  With the
successful running of B factories, high precision data on the rare
hadronic B decay modes such as $B\to \pi\pi, \pi K$
~\cite{Aubert:2002jm,Aubert:2002jb,Aubert:2002jj,Casey:2002yd,Abe:2001nq,Cronin-Hennessy:2000kg}
have been obtained, which provide us good opportunities to extract
the weak phase angle $\g$, to test the theoretical approaches for
evaluating the hadronic transition matrix elements,  and to
explore new physics beyond the SM.

 The recently proposed methods such as
QCD Factorization ~\cite{Beneke:1999br,Beneke:2000ry} and pQCD
approach ~\cite{Keum:2000ph,Keum:2000wi} have been extensively
discussed.  From those methods, useful information of weak phase
angles such as $\g$ can be
extracted\cite{Beneke:2001ev,Keum:2002ri}.

On the other hand, methods   based on flavor isospin and SU(3)
symmetries are still helpful and important
~\cite{zeppenfeld:1981ex,Savage:1989jx,Gronau:1995hm,He:1998rq,Paz:2002ev}
. The advantage of this kind of approaches is obvious that they
are model independent and more convenient in studying the
interference between weak and strong phases.  Recently the flavor
isospin and SU(3) symmetries in charmless B decays are studied by
using global fits to the experiment data
~\cite{Zhou:2000hg,He:2000ys,Fu:2002nr}.  In a general isospin
decomposition, there exist a lot of independent free parameters.
But using the flavor isospin and SU(3) symmetries, the number of
parameters can be greatly reduced and the method of global fit
becomes applicable.
Through direct fit, the isospin or SU(3) invariant amplitudes as
well as the corresponding strong phases can be extracted with a
reasonable precision. Our early results \cite{Zhou:2000hg} have
already indicated some unexpected large isospin amplitudes and
strong phases .  The fitted amplitudes and strong phases can also
provide useful information for the weak phase $\g$
~\cite{He:2000ys}. However unlike isospin symmetry, the flavor
SU(3) symmetry is known to be broken down sizably
\cite{Gronau:1998fn,Gronau:2000pk}.  The ways of introducing SU(3)
breaking may have significant influence on the final results. In
the usual considerations, the main effects of SU(3) breaking are
often taken into accounted only in the amplitudes.  To be more
general, the study of SU(3) breaking including strong phases is
necessary, which can be significant and bring a consistent
explanation to the present data\cite{WZ}. Of particular, their
effects can lead to different predictions on direct CP
violation\cite{WZ}.

 {\bf Isospin Relation}. Let us take decays $B\to \pi\pi$ as an example, the final states
of $\pi\pi$ have isospin of $2$ and $0$. The isospin amplitudes
$A_2$ and $A_0$ are defined as follows \be A_{2} &\equiv&
\bra{\pi\pi,I=2} H^{3/2}_{eff} \ket{B} =\lambda_{u}
a^{u}_{2}\exp{i\d^{u}_{2}}+\lambda_{c} a^{c}_{2}\exp{i\d^{c}_{2}},
\nn A_{0} &\equiv& \bra{\pi,\pi,I=0} H^{1/2}_{eff} \ket{B}
=\lambda_{u} a^{u}_{0}\exp{i\d^{u}_{0}}+\lambda_{c}
a^{c}_{0}\exp{i\d^{c}_{0}}, \ee where $a^{q}_{I}, (q=u,c \mbox{
and } I=2, 0)$ are the amplitudes associated with
$\lambda_{q}=V_{qb}V^{*}_{qd}$.
The isospin structure of effective Hermiltonian leads to the
following relations between two isospin amplitudes
\be\label{relation} \frac{a^c_2}{a^u_2} \equiv
R_{EW}=\frac{3}{2}\cdot \frac{C_9+C_{10}}{C_1+C_2+C_9+C_{10}}. \ee
%here the definition of $R_{EW}$ is slightly different from the one
%in Refs.\cite{Neubert:1998pt,Paz:2002ev}
%in which the CKM matrix elements have been included. The advantage of defining $R_{EW}$ in
%the above way is that it is directly connected to the one in $B\to \pi K$ through SU(3) symmetry.
Taking the Wilson coefficients at $\m=m_{b}$, one has $C_1=1.144,
C_2=-0.308, C_9=-1.28 \a$ and  $C_{10}=0.328 \a$.  Thus \be
R_{EW}=-1.25\times 10^{-2}, \ \ \mbox{and} \ \
\d^{c}_{2}=\d^{u}_{2}.  \ee

Note that the relation is obtained without the knowledge of the
matrix element $\bra{I=2} \order{3/2}\ket{B}$. It can not be
affected by the final state inelastic rescattering processes with
lower isospin as it is only related to the highest isospin
component $\D I=3/2$. Furthermore, the value of $R_{EW}$ is the
ratio between the electroweak penguin and tree diagrams. It is
then sensitive to new physics effects beyond the SM in electroweak
penguin sector. The new physics effects on $R_{EW}$ have been
discussed in
Refs.\cite{He:1999az,Grossman:1999av,Ghosh:2002jp,Xiao:2002mr}, it
seems quite sensitive to several new physics models.  A precise
determination of $R_{EW}$ from experiments may be helpful to
single out possible new physics or study flavor symmetry breaking
in charmless B decays.  To describe the possibility that the value
of $R_{EW}$ extracted from experiments could be different from the
SM calculations, we introduce a factor $\k$ as follows
\be R^{exp}_{EW}=\k \cdot R_{EW} \simeq -0.0125\cdot \k, \qquad
\quad k = 1 \quad \mbox{in} \quad \mbox{SM}
 \ee
 where $R^{exp}_{EW}$
stands for its value extracted from experiments and obviously
$\k=1$ in SM.

{\bf SU(3) Analysis}. In flavor $SU(3)$ limit, the decay
amplitudes for $B\to \pi\pi$ and $B\to \pi K$ is directly
connected
 \be\label{SU3relation}
a^{u}_{0}\exp{i\d^{u}_{0}}= a^{u}_{1/2}\exp{i\d^{u}_{1/2}}, \quad
a^{c}_{0}\exp{i\d^{c}_{0}}= a^{c}_{1/2}\exp{i\d^{c}_{1/2}}, \quad
a^{u}_{2}\exp{i\d^{u}_{2}}= a^{u}_{3/2}\exp{i\d^{u}_{3/2}}, \quad
a^{c}_{2}\exp{i\d^{c}_{2}}= a^{c}_{3/2}\exp{i\d^{c}_{3/2}}.
 \ee
If these relations are adopted, the number of free parameters is
reduced to be nine.
>From Eq. (\ref{relation}) and the above relation, one finds that
$\frac{a^{c}_{3/2}}{a^{u}_{3/2}}=\frac{a^{c}_{2}}{a^{u}_{2}}=R_{EW}$.
Thus the highest isospin amplitudes for the $B\to \pi K$ decays
satisfy the same relation as the one in the $B\to \pi\pi$ decay.
%%\marker{9}
When SU(3) breaking effects are considered, the above relations
have to be modified. At present stage, it is not very clear how to
describe the SU(3) breaking effects. a widely used approach is
introducing a breaking factor $\xi$ which characterizes the ratio
between $B\to \pi K$ and $\pi\pi$ decay amplitudes, i.e.,
%
%\be\label{simpleSU3brk}
$ a^{u(c)}_{1/2} = \xi a^{u(c)}_{0} , \ \ \ a^{u(c)}_{3/2} = \xi
a^{u(c)}_{2}, $
%\ee
%
but their strong phases are assumed to remain satisfying the SU(3)
relations
\be\label{EqualPhase} \delta^{u(c)}_{1/2} = \delta^{u(c)}_{0} , \
\ \ \delta^{u(c)}_{3/2} = \delta^{u(c)}_{2}. \ee
Typically $\xi=f_K/f_\pi\simeq 1.23$ with $f_\pi$ and $f_K$ being
the pion and kaon meson decay constants, which comes from the
naive factorization calculations.  It is easy to see that this
pattern of SU(3) breaking is a quite special one.  The value of
$\xi$ is highly model dependent. It can only serve as an order of
magnitude estimation and it is even not clear whether a single
factor can be applied to all the isospin amplitudes. The equal
strong phase assumption implies that the SU(3) breaking effects on
strong phase are all ignored, which may be far away from the
reality.  In a more general case, all the strong phases could be
different when SU(3) is broken down. The breaking effects on
strong phases may have significant effects on the prediction for
the direct CP violations in those decay modes.

%\marker{p20}%MARKER%
{\bf SU(3) Symmetry Breaking Effects of Strong Phases}. To
describe the possible violations of relations in eq.
(\ref{EqualPhase}) or the SU(3) breaking effects on strong phases,
we may introduce the following phase differences $\D^{q}_{I}
(q=u,c$ and $I=3/2, 1/2 )$: \be
\delta^{q}_{0}=\delta^{q}_{1/2}+\D^{q}_{1/2}, \ \ \
\delta^{q}_{2}= \delta^{q}_{3/2} + \D^{q}_{3/2} \ \ \  \ \
(q=u,c). \ee On the other hand, the SU(3) breaking effects in
amplitudes may also be given in a more general way
 \be\label{nosimpleSU3brk}
 a^{q}_{1/2} = \xi^q a^{u(c)}_{0} , \ \ \ a^{q}_{3/2} = \xi^q    a^{q}_{2} \ \ \  \ \ (q=u,c)
\ee The SU(3) limit corresponds to the case that all $\D^{q}_{I}$
vanish and $\xi^q =1$.

However, as it can be seen  in table \ref{simpleFit}, the global
fit to the latest data show that for $\xi=1$ or $1.23$ the best
fitted value of $\kappa$ is quite large, around 10. the fits with
different values of $\gamma$ show that the such a large  $\kappa$
insensitive to $\gamma$. this results is closely related to the
observed large branching ration of $B\to \pi^{0} K^{0}$ and $B\to
\pi^{0} \pi^{0}$.
\begin{table}[htb]%label{}
\caption{global fit of isospin amplitudes and strong phases in
charmless B decays with $\gamma=60^\circ$}
\begin{center}
\begin{ruledtabular}
\begin{tabular}{ccccc}
parameter & value(a) &value(b)&value(c)&value(d)\\\hline
 $a^u_{1/2}$&$     517.0^{+        81.5}_{       -80.6}$&$401.5^{+       125.1}_{      -205.2}$&$       293.8^{+        58.8}_{       -55.9}$&$       415.0^{+        77.8}_{       -77.8}$\\
 $\d^u_{1/2}$&$     2.42^{+         0.3}_{        -0.2}$&$    1.22^{+         0.3}_{        -1.5}$&$         0.7^{+         0.5}_{        -0.3}$&$         0.6^{+         0.4}_{        -0.3}$\\
 $a^c_{1/2}$&$     0.85^{+         2.9}_{        -2.9}$&$      -0.28^{+         2.9}_{        -2.8}$&$      -2.62^{+        2.46}_{       -1.97}$&$      1.18^{+        1.24}_{       -0.36}$\\
 $a^u_{3/2}$&$     536.8^{+        38.6}_{       -41.8}$&$       667.2^{+        48.1}_{       -51.8}$&$       432.4^{+        48.8}_{       -51.7}$&$       545.9^{+        51.4}_{       -54.6}$\\
 $\d^u_{3/2}$&$     3.09^{+         0.3}_{        -0.3}$&$      0.01^{+         1.2}_{        -0.3}$&$       1.43^{+         0.1}_{        -0.1}$&$       1.43^{+         0.1}_{        -0.1}$\\
 $b^c_{1/2}$&$    -141.0^{+         4.2}_{        -4.2}$&$      -148.0^{+         4.2}_{        -4.1}$&$      -132.1^{+        15.5}_{       -10.9}$&$      -127.3^{+        16.6}_{       -12.1}$\\
 $\d'^u_{1/2}$&$       2.8^{+         0.4}_{        -0.5}$&$       -0.28^{+         1.0}_{        -0.4}$&$        -0.1^{+         0.2}_{        -0.2}$&$        -0.2^{+         0.2}_{        -0.2}$\\
$\xi$ & 1.0(fix) & 1.23(fix) & 1.0(fix) &1.23(fix)\\
$\k$  & 1.0(fix) & 1.0 (fix) &  $12.0^{+         5.3}_{
-4.4}$&$10.7^{+         3.6}_{        -3.2}$\\\hline
$\chi^{2}_{min}$& 5.8 & 9.2 & 0.61 &0.85
\end{tabular}
\end{ruledtabular}
\label{simpleFit}
\end{center}
\end{table}
The value of $\k$ is sensitive to the contributions from
electroweak penguin diagrams.  Since many new physics models can
give significant corrections to this sector, it may be helpful to
study new physics effects on $\k$.  However, to explore any new
physics effects and arrive at a definitive conclusion for the
existence of new physics from the hadronic decays, it is necessary
to check all the theoretical assumptions and make the most general
considerations.  It is noted that the above results are obtained
by assuming SU(3) symmetry with its breaking only in amplitudes.
Therefore, we need first examine the above results to a more
general case of SU(3) symmetry breaking before claiming any
possible new physics signals. Within the framework of the SM, we
found   that the breaking effects of flavor SU(3) symmetry could
be considerable.

We then explore the parameter space of those $SU(3)$ breaking
factors and found that for some typical non-zero values of the
phase shift $\Delta^{q}_{I}$ the best fit of $\kappa$ does restore
to it's values in SM, i.e. close to unity. For example, in cases
of $\D^{u}_{1/2}=+\pi/6$, $\D^{c}_{1/2}=+\pi/6, +\pi/3$ and
$\D^{c}_{3/2}= +\pi/6, +\pi/3$, the best fitted values of $\k$ are
around 1.5 with the minimal $\chi^2_{min}\leq 4$. The results is
similar for other values of $\gamma$ as long as $\gamma
<120^{\circ}$.

{\bf Direct CP Violation}. From the fit result, the corresponding
direct CP violation can also be obtained.  The best fitted direct
CP violation for example, in case (c) is given by
\begin{eqnarray}
& & A_{CP}(\pi^{+}\pi^{-})  \simeq 0.3 \quad  \qquad  A_{CP}(\pi^{0}\pi^{0}) \simeq 0.4 \nonumber \\
 & & A_{CP}(\pi^{+} K^-) \simeq -0.1 \quad \qquad  A_{CP}(\pi^{0}\bar K^{0}) \simeq -0.1 \\
 & &  A_{CP}(\pi^{0} K^-) \simeq -0.0  \quad \qquad A_{CP}(\pi^{-} \bar K^0) \simeq 0.1 \nonumber
\end{eqnarray}
In cases of $\Delta^{u}_{1/2}=+\pi/6$, $\Delta^{c}_{1/2}=+\pi/6,
+\pi/3$ and $\Delta^{c}_{3/2}= +\pi/6, +\pi/3$, the best fitted
values of $\kappa$ are around 1.5 with the minimal
$\chi^2_{min}\leq 4$.  The direct CP violation for
$\Delta^{u}_{1/2}=+\pi/6(\Delta^{c}_{1/2}= +\pi/6)$ is as follows.
\begin{eqnarray}
 & & A_{CP}(\pi^{+}\pi^{-}) \simeq 0.1(  0.5 ), \qquad \quad  A_{CP}(\pi^{0}\pi^{0}) \simeq  0.5(0.2 ), \nonumber
 \\
 & &  A_{CP}(\pi^{+} K^-) \simeq -0.1(-0.1), \qquad \quad A_{CP}(\pi^{0}\bar K^{0}) \simeq  -0.2(-0.1 ),  \\
 & &  A_{CP}(\pi^{0} K^-) \simeq -0.1(-0.0), \qquad \quad A_{CP}(\pi^{-} \bar K^0) \simeq  0.1(0.1). \nonumber
\end{eqnarray}
Compared with the ones with SU(3) syemmetry case, the predicted
values of direct CP violation can be quite different.

 {\bf Summary II}. In the case of SU(3) limits and also the case with
SU(3) breaking only in amplitudes, the fitting results lead to an
unexpected large ratio between two isospin amplitudes
$a^{c}_{3/2}/a^{u}_{3/2}$, which is about an order of magnitude
larger than the SM prediction.  The results are found to be
insensitive to the weak phase $\gamma$.  By including SU(3)
breaking effects on the strong phases, one is able to obtain a
consistent fit to the current data within the SM, which implies
that the SU(3) breaking effect on strong phases may play an
important role in understanding the observed charmless hadronic B
decay modes $B\to \pi \pi$ and $\pi K$. It is possible to test
those breaking effects in the near future from more precise
measurements of direct CP violation in B factories.
\\

{\bf Acknowledgement}: This work is supported in part by key
project of the Chinese Academy of Sciences and National Natural
Science Foundation of China (NSFC). YLW would like to thank  C.W.
Kim for his kind invitation and the local organization committee
chaired by E.J. Chun for the excellent organization on the
ICFP2003 held at KIAS, Korea. He also wants to acknowledge the
associate scheme at the Abdus Salam I.C.T.P., Italy, where the
proceedings paper was prepared. YFZ acknowledges the support by
Alexander von Humboldt Foundation.


\begin{thebibliography}{27}
%\begin{references}

\expandafter\ifx\csname
natexlab\endcsname\relax\def\natexlab#1{#1}\fi
\expandafter\ifx\csname bibnamefont\endcsname\relax
  \def\bibnamefont#1{#1}\fi
\expandafter\ifx\csname bibfnamefont\endcsname\relax
  \def\bibfnamefont#1{#1}\fi
\expandafter\ifx\csname citenamefont\endcsname\relax
  \def\citenamefont#1{#1}\fi
\expandafter\ifx\csname url\endcsname\relax
  \def\url#1{\texttt{#1}}\fi
\expandafter\ifx\csname
urlprefix\endcsname\relax\def\urlprefix{URL }\fi
\providecommand{\bibinfo}[2]{#2}
\providecommand{\eprint}[2][]{\url{#2}}

\bibitem{HF} H. Fritzsch, theory summary talk in this proceedings.
\bibitem{KK} K. Kleinknecht, experiment summary talk in this
proceedings.
\bibitem{NA48} J.R. Batley, et.al., NA48 collaboration, Phys.Lett. {\bf B 544},
97 (2002).
\bibitem{KTeV} KTeV Collaboration: A. Alavi-Harati, et al, Phys.Rev. {\bf D 67},
   012005 (2003).
\bibitem{ylw} Y.L. Wu, Phys. Rev. {\bf D 64}, 016001 (2001); More
references can be found in: Y.L. Wu, Plenary talk at International
Conference on Flavor Physics (ICFP 2001), published in {\bf Flavor
physics} 217-236 (2002) (World Scientfic Pub. Co.),
hep-ph/0108155.
\bibitem[{\citenamefont{Kobayashi and Maskawa}(1973)}]{kobayashi:1973fv}
\bibinfo{author}{\bibfnamefont{M.}~\bibnamefont{Kobayashi}} \bibnamefont{and}
  \bibinfo{author}{\bibfnamefont{T.}~\bibnamefont{Maskawa}},
  \bibinfo{journal}{Prog. Theor. Phys.} \textbf{\bibinfo{volume}{49}},
  \bibinfo{pages}{652} (\bibinfo{year}{1973}).
\bibitem{W0} Y. L. Wu, Mod. Phys. Lett. A {\bf 8}, 819 (1993).
\bibitem{W1} W. Y. Wang, Y. L. Wu and Y. A. Yan, Int. J. Mod. Phys. {\bf A
15}, 1817 (2000).
\bibitem{W2} Y. A. Yan, Y. L. Wu and W. Y. Wang, Int. J. Mod. Phys. {\bf A
15}, 2375 (2000).
\bibitem{W3} W. Y. Wang and Y. L. Wu, Int. J. Mod. Phys. {\bf A 16}, 377 (2001).
\bibitem{W4} Y. L. Wu and Y. A. Yan, Int. J. Mod. Phys. {\bf A 16}, 285 (2001).
\bibitem{W5} W. Y. Wang and Y. L. Wu, Phys. Lett. {\bf B 515}, 57 (2001).
\bibitem{W6} W. Y. Wang and Y. L. Wu, Phys. Lett. {\bf B 519}, 219 (2001).
\bibitem{W7} W. Y. Wang and Y. L. Wu, M. Zhong, Phys. Rev. {\bf D 67}, 014024 (2003).
\bibitem{W8} M. Zhong, Y. L. Wu and W. Y. Wang, Int. J. Mod. Phys. {\bf A 18}, 1959 (2003).
\bibitem{W9} W. Y. Wang, Y. L. Wu and M. Zhong  J. Phys. {\bf G
29}, 2743 (2003).
\bibitem{W10} Y.L. Wu, Y.A. Yan, M. Zhong, Y.B. Zuo and W.Y. Wang,
     Mod. Phys. Lett. A {\bf 18}, 1303 (2003).
\bibitem{W11} W.Y. Wang, Y.L. Wu, Y.A. Yan, M. Zhong and Y.B. Zuo,
hep-ph/0310266.
\bibitem{HQS} N. Isgur, M. Wise, Phys. Lett. {\bf B 232}, 113 (1989);
   {\bf B 237}, 527 (1990); {\bf B 206}, 681 (1988).
\bibitem{HQL} M. B. Voloshin, M. A. Shifman, Sov. J. Nucl. Phys.
{\bf 45}, 292 (1987); {\bf 47}, 199 (1988); \\ E. Eichten, B.
Hill, Phys. Lett. {\bf B 234}, 511 (1990); E. Eichten, Nucl. Phys.
Proc. Suppl. {\bf B 4}, 170 (1988).
\bibitem{HQET} H. Georgi, Phys. Lett. {\bf B 240}, 447 (1990).
\bibitem{MR} T. Mannel, W. Roberts, Z. Ryzak, Nucl. Phys. {\bf B 368}, 204 (1992).
\bibitem{HQA} For review see e.g., M. Neubert, Phys. Rept. {\bf 245}, 259
(1994).
\bibitem{ac} A. Czarnecki, Phys. Rev. Lett. {\bf 22}, 4124 (1996).
\bibitem{am} A. F. Falk, M. Neubert, Phys. Rev. {\bf D 47}, 2965 (1993).
\bibitem{mc} M. Neubert, C. Sachrajda, Nucl. Phys. {\bf B 483}, 339 (1997);
     M. Neubert, hep-ph/9707217; and references therein.
\bibitem{mjch} M. S. Baek, J. Lee, C. Liu, H. S. Song, Phys. Rev. {\bf
     D 57}, 4091 (1998).
%\bibitem{W2} Y. A. Yan, Y. L. Wu, W. Y. Wang, Int. J. Mod. Phys. {\bf A 15}, 2735 (2000).
%\bibitem{W4} Y. L. Wu, Y. A. Yan, Int. J. Mod. Phys. {\bf A 16}, 285 (2001).
\bibitem{mmm} M. Luke, M. J. Savage, M. B. Wise, Phys. Lett. {\bf B 345}, 301 (1995).
\bibitem{mmmb} M. Lu, M. Luke, M. J. Savage, B. H. Smith, Phys. Rev. {\bf D 55}, 2827
     (1997).
%\bibitem{W3} W. Y. Wang, Y. L. Wu, Int. J. Mod. Phys. A {\bf 16}, 377 (2001).
\bibitem{PDG} K. Hagiwara et al., Phys. Rev. {\bf D 66}, 010001 (2002);
     and references therein.
\bibitem{LW} L. Wolfenstein, Phys. Rev. Lett. {\bf 51} (1983) 1945.
\bibitem{gzmy} G. Burdman, Z. Ligeti, M. Neubert, Y. Nir, Phys. Rev. {\bf D 49},
     2331 (1994).
\bibitem{hly} C. S. Huang, C. Liu, C. T. Yan, Phys. Rev. {\bf D 62}, 054019 (2000).
%\bibitem{pub5} W. Y. Wang, Y. L. Wu, Phys. Lett. B {\bf 515}, 57 (2001).
%\bibitem{pub6} W. Y. Wang, Y. L. Wu, Phys. Lett. B {\bf 519}, 219 (2001).
%\bibitem{nloBpi} W.Y. Wang, Y.L. Wu, M. Zhong, J. Phys. G {\bf 29}, 2743
%     (2003).
\bibitem[{\citenamefont{Aubert et~al.}(2002{\natexlab{a}})}]{Aubert:2002jm}
\bibinfo{author}{\bibfnamefont{B.}~\bibnamefont{Aubert}} \bibnamefont{et~al.}
  (\bibinfo{collaboration}{BABAR}) (\bibinfo{year}{2002}{\natexlab{a}}),
  \eprint[http://arXiv.org/abs]{hep-ex/0207065}.

\bibitem[{\citenamefont{Aubert et~al.}(2002{\natexlab{b}})}]{Aubert:2002jb}
\bibinfo{author}{\bibfnamefont{B.}~\bibnamefont{Aubert}} \bibnamefont{et~al.}
  (\bibinfo{collaboration}{BABAR}) (\bibinfo{year}{2002}{\natexlab{b}}),
  \eprint[http://arXiv.org/abs]{hep-ex/0207055}.

\bibitem[{\citenamefont{Aubert et~al.}(2002{\natexlab{c}})}]{Aubert:2002jj}
\bibinfo{author}{\bibfnamefont{B.}~\bibnamefont{Aubert}} \bibnamefont{et~al.}
  (\bibinfo{collaboration}{BABAR}) (\bibinfo{year}{2002}{\natexlab{c}}),
  \eprint[http://arXiv.org/abs]{hep-ex/0207063}.

\bibitem[{\citenamefont{Casey}(2002)}]{Casey:2002yd}
\bibinfo{author}{\bibfnamefont{B.~C.~K.} \bibnamefont{Casey}}
  (\bibinfo{collaboration}{Belle}) (\bibinfo{year}{2002}),
  \eprint[http://arXiv.org/abs]{hep-ex/0207090}.

\bibitem[{\citenamefont{Abe et~al.}(2001)}]{Abe:2001nq}
\bibinfo{author}{\bibfnamefont{K.}~\bibnamefont{Abe}} \bibnamefont{et~al.}
  (\bibinfo{collaboration}{BELLE}), \bibinfo{journal}{Phys. Rev. Lett.}
  \textbf{\bibinfo{volume}{87}}, \bibinfo{pages}{101801}
  (\bibinfo{year}{2001}), \eprint[http://arXiv.org/abs]{hep-ex/0104030}.

\bibitem[{\citenamefont{Cronin-Hennessy et~al.}(2000)}]{Cronin-Hennessy:2000kg}
\bibinfo{author}{\bibfnamefont{D.}~\bibnamefont{Cronin-Hennessy}}
  \bibnamefont{et~al.} (\bibinfo{collaboration}{CLEO}), \bibinfo{journal}{Phys.
  Rev. Lett.} \textbf{\bibinfo{volume}{85}}, \bibinfo{pages}{515}
  (\bibinfo{year}{2000}).

\bibitem[{\citenamefont{Beneke et~al.}(1999)\citenamefont{Beneke, Buchalla,
  Neubert, and Sachrajda}}]{Beneke:1999br}
\bibinfo{author}{\bibfnamefont{M.}~\bibnamefont{Beneke}},
  \bibinfo{author}{\bibfnamefont{G.}~\bibnamefont{Buchalla}},
  \bibinfo{author}{\bibfnamefont{M.}~\bibnamefont{Neubert}}, \bibnamefont{and}
  \bibinfo{author}{\bibfnamefont{C.~T.} \bibnamefont{Sachrajda}},
  \bibinfo{journal}{Phys. Rev. Lett.} \textbf{\bibinfo{volume}{83}},
  \bibinfo{pages}{1914} (\bibinfo{year}{1999}), \eprint{hep-ph/9905312}.

\bibitem[{\citenamefont{Beneke et~al.}(2000)\citenamefont{Beneke, Buchalla,
  Neubert, and Sachrajda}}]{Beneke:2000ry}
\bibinfo{author}{\bibfnamefont{M.}~\bibnamefont{Beneke}},
  \bibinfo{author}{\bibfnamefont{G.}~\bibnamefont{Buchalla}},
  \bibinfo{author}{\bibfnamefont{M.}~\bibnamefont{Neubert}}, \bibnamefont{and}
  \bibinfo{author}{\bibfnamefont{C.~T.} \bibnamefont{Sachrajda}},
  \bibinfo{journal}{Nucl. Phys.} \textbf{\bibinfo{volume}{B591}},
  \bibinfo{pages}{313} (\bibinfo{year}{2000}), \eprint{hep-ph/0006124}.

\bibitem[{\citenamefont{Keum et~al.}(2001{\natexlab{a}})\citenamefont{Keum, Li,
  and Sanda}}]{Keum:2000ph}
\bibinfo{author}{\bibfnamefont{Y.-Y.} \bibnamefont{Keum}},
  \bibinfo{author}{\bibfnamefont{H.-n.} \bibnamefont{Li}}, \bibnamefont{and}
  \bibinfo{author}{\bibfnamefont{A.~I.} \bibnamefont{Sanda}},
  \bibinfo{journal}{Phys. Lett.} \textbf{\bibinfo{volume}{B504}},
  \bibinfo{pages}{6} (\bibinfo{year}{2001}{\natexlab{a}}),
  \eprint{hep-ph/0004004}.

\bibitem[{\citenamefont{Keum et~al.}(2001{\natexlab{b}})\citenamefont{Keum, Li,
  and Sanda}}]{Keum:2000wi}
\bibinfo{author}{\bibfnamefont{Y.~Y.} \bibnamefont{Keum}},
  \bibinfo{author}{\bibfnamefont{H.-N.} \bibnamefont{Li}}, \bibnamefont{and}
  \bibinfo{author}{\bibfnamefont{A.~I.} \bibnamefont{Sanda}},
  \bibinfo{journal}{Phys. Rev.} \textbf{\bibinfo{volume}{D63}},
  \bibinfo{pages}{054008} (\bibinfo{year}{2001}{\natexlab{b}}),
  \eprint[http://arXiv.org/abs]{hep-ph/0004173}.

\bibitem[{\citenamefont{Beneke et~al.}(2001)\citenamefont{Beneke, Buchalla,
  Neubert, and Sachrajda}}]{Beneke:2001ev}
\bibinfo{author}{\bibfnamefont{M.}~\bibnamefont{Beneke}},
  \bibinfo{author}{\bibfnamefont{G.}~\bibnamefont{Buchalla}},
  \bibinfo{author}{\bibfnamefont{M.}~\bibnamefont{Neubert}}, \bibnamefont{and}
  \bibinfo{author}{\bibfnamefont{C.~T.} \bibnamefont{Sachrajda}},
  \bibinfo{journal}{Nucl. Phys.} \textbf{\bibinfo{volume}{B606}},
  \bibinfo{pages}{245} (\bibinfo{year}{2001}), \eprint{hep-ph/0104110}.

\bibitem[{\citenamefont{Keum}(2002)}]{Keum:2002ri}
\bibinfo{author}{\bibfnamefont{Y.-Y.} \bibnamefont{Keum}}
  (\bibinfo{year}{2002}), \eprint[http://arXiv.org/abs]{hep-ph/0209208}.

\bibitem[{\citenamefont{Zeppenfeld}(1981)}]{zeppenfeld:1981ex}
\bibinfo{author}{\bibfnamefont{D.}~\bibnamefont{Zeppenfeld}},
  \bibinfo{journal}{Zeit. Phys.} \textbf{\bibinfo{volume}{C8}},
  \bibinfo{pages}{77} (\bibinfo{year}{1981}).

\bibitem[{\citenamefont{Savage and Wise}(1989)}]{Savage:1989jx}
\bibinfo{author}{\bibfnamefont{M.~J.} \bibnamefont{Savage}} \bibnamefont{and}
  \bibinfo{author}{\bibfnamefont{M.~B.} \bibnamefont{Wise}},
  \bibinfo{journal}{Nucl. Phys.} \textbf{\bibinfo{volume}{B326}},
  \bibinfo{pages}{15} (\bibinfo{year}{1989}).

\bibitem[{\citenamefont{Gronau et~al.}(1995)\citenamefont{Gronau, Hernandez,
  and London}}]{Gronau:1995hm}
\bibinfo{author}{\bibfnamefont{M.}~\bibnamefont{Gronau}},
  \bibinfo{author}{\bibfnamefont{O.~F.} \bibnamefont{Hernandez}},
  \bibnamefont{and} \bibinfo{author}{\bibnamefont{London}},
  \bibinfo{journal}{Phys. Rev.} \textbf{\bibinfo{volume}{D52}},
  \bibinfo{pages}{6356} (\bibinfo{year}{1995}), \eprint{hep-ph/9504326}.

\bibitem[{\citenamefont{He}(1999)}]{He:1998rq}
\bibinfo{author}{\bibfnamefont{X.-G.} \bibnamefont{He}}, \bibinfo{journal}{Eur.
  Phys. J.} \textbf{\bibinfo{volume}{C9}}, \bibinfo{pages}{443}
  (\bibinfo{year}{1999}), \eprint{hep-ph/9810397}.

\bibitem[{\citenamefont{Paz}(2002)}]{Paz:2002ev}
\bibinfo{author}{\bibfnamefont{G.}~\bibnamefont{Paz}} (\bibinfo{year}{2002}),
  \eprint[http://arXiv.org/abs]{hep-ph/0206312}.

\bibitem[{\citenamefont{Zhou et~al.}(2001)\citenamefont{Zhou, Wu, Ng, and
  Geng}}]{Zhou:2000hg}
\bibinfo{author}{\bibfnamefont{Y.~F.} \bibnamefont{Zhou}},
  \bibinfo{author}{\bibfnamefont{Y.~L.} \bibnamefont{Wu}},
  \bibinfo{author}{\bibfnamefont{J.~N.} \bibnamefont{Ng}}, \bibnamefont{and}
  \bibinfo{author}{\bibfnamefont{C.~Q.} \bibnamefont{Geng}},
  \bibinfo{journal}{Phys. Rev.} \textbf{\bibinfo{volume}{D63}},
  \bibinfo{pages}{054011} (\bibinfo{year}{2001}),
  \eprint[http://arXiv.org/abs]{hep-ph/0006225}.

\bibitem[{\citenamefont{He et~al.}(2001)}]{He:2000ys}
\bibinfo{author}{\bibfnamefont{X.~G.} \bibnamefont{He}} \bibnamefont{et~al.},
  \bibinfo{journal}{Phys. Rev.} \textbf{\bibinfo{volume}{D64}},
  \bibinfo{pages}{034002} (\bibinfo{year}{2001}), \eprint{hep-ph/0011337}.

\bibitem[{\citenamefont{Fu et~al.}(2002)\citenamefont{Fu, He, Hsiao, and
  Shi}}]{Fu:2002nr}
\bibinfo{author}{\bibfnamefont{H.-K.} \bibnamefont{Fu}},
  \bibinfo{author}{\bibfnamefont{X.-G.} \bibnamefont{He}},
  \bibinfo{author}{\bibfnamefont{Y.-K.} \bibnamefont{Hsiao}}, \bibnamefont{and}
  \bibinfo{author}{\bibfnamefont{J.-Q.} \bibnamefont{Shi}}
  (\bibinfo{year}{2002}), \eprint[http://arXiv.org/abs]{hep-ph/0206199}.

\bibitem[{\citenamefont{Gronau et~al.}(1999)\citenamefont{Gronau, Pirjol, and
  Yan}}]{Gronau:1998fn}
\bibinfo{author}{\bibfnamefont{M.}~\bibnamefont{Gronau}},
  \bibinfo{author}{\bibfnamefont{D.}~\bibnamefont{Pirjol}}, \bibnamefont{and}
  \bibinfo{author}{\bibfnamefont{T.-M.} \bibnamefont{Yan}},
  \bibinfo{journal}{Phys. Rev.} \textbf{\bibinfo{volume}{D60}},
  \bibinfo{pages}{034021} (\bibinfo{year}{1999}),
  \eprint[http://arXiv.org/abs]{hep-ph/9810482}.

\bibitem[{\citenamefont{Gronau and Pirjol}(2000)}]{Gronau:2000pk}
\bibinfo{author}{\bibfnamefont{M.}~\bibnamefont{Gronau}} \bibnamefont{and}
  \bibinfo{author}{\bibfnamefont{D.}~\bibnamefont{Pirjol}},
  \bibinfo{journal}{Phys. Rev.} \textbf{\bibinfo{volume}{D62}},
  \bibinfo{pages}{077301} (\bibinfo{year}{2000}),
  \eprint[http://arXiv.org/abs]{hep-ph/0004007}.

\bibitem{WZ} Y.L. Wu, Y.F. Zhou, to be published in EPJC,
hep-ph/0210367.

\bibitem[{\citenamefont{He et~al.}(2000)\citenamefont{He, Hsueh, and
  Shi}}]{He:1999az}
\bibinfo{author}{\bibfnamefont{X.-G.} \bibnamefont{He}},
  \bibinfo{author}{\bibfnamefont{C.-L.} \bibnamefont{Hsueh}}, \bibnamefont{and}
  \bibinfo{author}{\bibfnamefont{J.-Q.} \bibnamefont{Shi}},
  \bibinfo{journal}{Phys. Rev. Lett.} \textbf{\bibinfo{volume}{84}},
  \bibinfo{pages}{18} (\bibinfo{year}{2000}),
  \eprint[http://arXiv.org/abs]{hep-ph/9905296}.

\bibitem[{\citenamefont{Grossman et~al.}(1999)\citenamefont{Grossman, Neubert,
  and Kagan}}]{Grossman:1999av}
\bibinfo{author}{\bibfnamefont{Y.}~\bibnamefont{Grossman}},
  \bibinfo{author}{\bibfnamefont{M.}~\bibnamefont{Neubert}}, \bibnamefont{and}
  \bibinfo{author}{\bibfnamefont{A.~L.} \bibnamefont{Kagan}},
  \bibinfo{journal}{JHEP} \textbf{\bibinfo{volume}{10}}, \bibinfo{pages}{029}
  (\bibinfo{year}{1999}), \eprint[http://arXiv.org/abs]{hep-ph/9909297}.

\bibitem[{\citenamefont{Ghosh et~al.}(2002)\citenamefont{Ghosh, He, Hsiao, and
  Shi}}]{Ghosh:2002jp}
\bibinfo{author}{\bibfnamefont{D.~K.} \bibnamefont{Ghosh}},
  \bibinfo{author}{\bibfnamefont{X.-G.} \bibnamefont{He}},
  \bibinfo{author}{\bibfnamefont{Y.-K.} \bibnamefont{Hsiao}}, \bibnamefont{and}
  \bibinfo{author}{\bibfnamefont{J.-Q.} \bibnamefont{Shi}}
  (\bibinfo{year}{2002}), \eprint[http://arXiv.org/abs]{hep-ph/0206186}.

\bibitem[{\citenamefont{Xiao et~al.}(2002)\citenamefont{Xiao, Chao, and
  Li}}]{Xiao:2002mr}
\bibinfo{author}{\bibfnamefont{Z.-j.} \bibnamefont{Xiao}},
  \bibinfo{author}{\bibfnamefont{K.-T.} \bibnamefont{Chao}}, \bibnamefont{and}
  \bibinfo{author}{\bibfnamefont{C.~S.} \bibnamefont{Li}},
  \bibinfo{journal}{Phys. Rev.} \textbf{\bibinfo{volume}{D65}},
  \bibinfo{pages}{114021} (\bibinfo{year}{2002}), \eprint{hep-ph/0204346}.

\end{thebibliography}
\end{document}